\documentclass[prb,twocolumn,showpacs,preprintnumbers]{revtex4}
\usepackage{amssymb}
\usepackage{bm}
\usepackage{amsmath}
\usepackage[dvips]{graphicx}
\usepackage{color}

\begin{document}

\title{Quasielastic neutron scattering from two dimensional antiferromagnets
at a finite temperature }
\author{A. Katanin}
\affiliation{Institute of Metal Physics, 620041, Ekaterinburg, Russia\\
Ural Federal University, 620002, Ekaterinburg, Russia}
\author{O. P. Sushkov}
\affiliation{School of Physics, University of New South Wales, Sydney 2052, Australia}

\begin{abstract}
We consider frequency dependence of the neutron scattering amplitude from a
two-dimensional quantum antiferromagnet. It is well known that the long
range order disappears at any finite temperature and hence the elastic
neutron scattering Bragg peak is transformed to the quasielastic neutron
scattering spectrum $\propto d\omega /\omega $. We show that the widely
known formula for the spectrum of an isotropic antiferromagnet derived by
Auerbach and Arovas~\cite{AA} should be supplemented by a logarithmic term
that changes the integrated intensity by two times. A similar formula for an
easy-plane magnet is very much different because of the
Berezinsky-Kosterlitz-Thouless physics. An external uniform magnetic field
switches smoothly the isotropic magnet to the easy-plane magnet. We
demonstrate that the quasielastic neutron scattering spectrum in the
crossover regime combines properties of both limiting cases. We also
consider a quantum antiferromagnet close to the O(3) quantum critical point
and show that in an external uniform magnetic field the intensity of elastic
(quasielastic) neutron scattering peak depends linearly and significantly on
the applied field.
\end{abstract}

\date{\today }
\maketitle

\section{Introduction}

The two-dimensional antiferromagnets have been studied
thoroughly in numerous works. Nevertheless, surprisingly, some basic
questions related to the elastic and quasielastic neutron scattering remain
unresolved. It is widely known that the (quasi-)two-dimensionality of
layered systems yields logarithmic corrections to the sublattice
magnetization, amplitude of magnon scattering \cite{Polyakov,Chub1} and spin
correlation functions\cite{AA,Chak,Chub}. Both, magnetic field or easy-plane
magnetic anisotropy weaken the above-mentioned divergent contributions and
simultaneously yield a Kosterlitz-Thouless behavior at low temperatures.
Studying neutron scattering amplitude in the presence of these factors is an
important problem, which is also relevant from the experimental point of
view.

Our interest to this problem has been stimulated by recent experimental
studies~\cite{Hinkov08,Haug09} of quasielastic neutron scattering from
underdoped YBa$_{2}$Cu$_{3}$O$_{6.45}$. This sample corresponds to the hole
doping level about 8.5\%. The paper~\cite{Hinkov08} reports a weak
quasielastic neutron scattering and hence indicates a quasistatic magnetic
ordering. This demonstrates proximity to a quantum critical point (QCP)
between magnetically disordered and magnetically ordered states. The paper~%
\cite{Haug09} reports that the quasielastic neutron scattering intensity
depends substantially on the applied uniform magnetic field. Earlier a
similar effect of enhancement of the quasielastic scattering in magnetic
field was observed in La$_{1.9}$Sr$_{0.1}$CuO$_{4}$, Ref.~\onlinecite{Lake02}. 
We believe
that the effect in both compounds is of the same origin. However, a
theoretical analysis in La$_{2-x}$Sr$_{x}$CuO$_{4}$ is much harder due to a
large degree of intrinsic disorder related to random Sr positions. The
observed magnetism is two-dimensional (2D), there are no indications for a
correlation in the direction perpendicular to CuO$_{2}$ planes. Furthermore,
it is incommensurate, mainly due to the contribution of charge degrees
of freedom, which certainly complicates a theoretical analysis of magnetic
properties, see Ref.~\onlinecite{Milstein08}.  Moreover, both
compounds, YBa$_{2}$Cu$_{3}$O$_{6.45}$ and La$_{1.9}$Sr$_{0.1}$CuO$_{4}$
 are superconducting at low enough temperatures.

In the present paper we simplify the problem and consider two dimensional
quantum antiferromagnets with ordered ground state described by the
Heisenberg model. The presented analysis can be also relevant
for the layered perovskite compounds, in particular K$_{2}$MnF$_{4}$ and K$%
_{2}$CuF$_{4}$ (Ref. \onlinecite{Joungh}).  We are interested in temperatures and
external magnetic fields that are much smaller than the Heisenberg exchange
interaction J. Therefore, it is quite natural to apply techniques of the
nonlinear $\sigma $-model that significantly simplifies calculations. It is
known for long time that the model describes the low-energy properties of the
collinear quantum magnets~\cite{Chak,Chub}.

A quantum magnet in a ground state spontaneously violates the continuous
symmetry of the Hamiltonian and this leads to the spontaneous magnetization
and, due to the Goldstone theorem, to gapless magnons. Neutron magnetic
scattering from the static magnetization leads to the elastic Bragg peak.
Due to the gapless excitation spectrum and due to the low dimensionality,
any finite temperature destroys the long range order and hence destroys the
elastic Bragg peak. The peak is transformed to the narrow quasielastic
spectrum. The first problem that we address in the present paper is the
derivation of this quasielastic spectrum.

The second problem that we address in the present paper is how the intensity
of elastic (quasielastic) neutron scattering depends on the applied uniform
magnetic field if the system is in the vicinity of a QCP. The obtained
result is generic and is independent of the specific mechanism of the
quantum phase transition. We believe that it explains the data obtained in
Refs.~\onlinecite{Haug09,Lake02}.

The structure of the paper is the following: In Section II we formulate
problems and present answers for quasielastic neutron scattering spectra 
at nonzero temperature.
In Section III we discuss the
dependence of elastic neutron scattering on applied magnetic field and
compare theory with experimental data.
Section IV presents simple heuristic derivations of
quasielastic neutron scattering spectra at nonzero temperature.
A rigorous derivation of quasielastic spectra based on the
Renormalization Group (RG) analysis is given in Section V. Section VI
presents our conclusions.

\section{Questions and Answers}

In the present Section we only formulate problems and present our results
for neutron scattering spectra/probabilities in various cases. Derivations
of these results are presented in sections IV and V.

The Lagrangian of the nonlinear O(3) $\sigma $-model describing a 2D
isotropic quantum antiferromagnet reads~\cite{Chak} 
\begin{equation}
\mathcal{L}=\frac{1}{2}\chi _{\perp }{\dot{\vec{n}}}^{2}-\frac{1}{2}\rho
_{s}({\bm{\nabla}}{\vec{n}})^{2}\ ,  \label{L1}
\end{equation}%
where $\chi _{\perp }$ is the magnetic susceptibility and $\rho _{s}$ is the
spin stiffness. The unit vector, ${\vec{n}}=(n_{x},n_{y},n_{z})$, $n^{2}=1$,
describes staggered spins. Here we consider \textquotedblleft quantum
renormalized\textquotedblright\ model with $\chi _{\perp }$ and $\rho _{s}$
being renormalized (observed) parameters of the ground state. We use the
choice of normalization (cutoff), such that $\langle n_{z}\rangle _{T=0}=1$.
The ground-state sublattice magnetization $\mu $ is considered as an
independent parameter of the theory~\cite{Ren}. Also, we consider real time
since our aim is the real excitation spectrum. Throughout the paper we set
the Planck's constant and the Boltzmann's constant equal to unity, $\hbar
=k_{B}=1$. Interaction with the incident neutron we take in the following
simplified form that is sufficient for our purposes 
\begin{equation}
\mathcal{L}_{int}=\psi _{\bm r}^{\dag }({\vec{\sigma}}\cdot {\vec{n}}({\bm r}%
))\psi _{\bm r}\ ,  \label{nint}
\end{equation}%
where ${\vec{\sigma}}$ is the Pauli matrix describing the neutron spin, and $%
\psi _{\bm r}$ is the wave function of the neutron. Since ${\vec{n}}$ is the
staggered magnetization, Eq.(\ref{nint}) implies that the momentum transfer
is shifted by the AF vector. We also introduce external magnetic field by
changing ${\dot{\vec{n}}\rightarrow \dot{\vec{n}}-}\overrightarrow{{B}}{%
\times }\overrightarrow{{n}}$.

At zero temperature the O(3) rotational symmetry is spontaneously broken and
the ground state magnetization is ${\vec n}_0=(0,0,1)$. Quantum fluctuations
around this ground state are, ${\vec n}={\vec n}_0+{\vec n}_{\perp}$, 
\begin{eqnarray}  \label{q1}
&&{\vec n}_{\perp}=\sum_{{\bm k},\lambda=x,y} {\vec e}_{\lambda}A_{\bm k}
\left(a_{{\bm k}\lambda}e^{-i\omega_{\bm k}t+i{\bm k}\cdot{\bm r}}+ a_{{\bm k%
}\lambda}^{\dag}e^{i\omega_{\bm k}t-i{\bm k}\cdot{\bm r}} \right)  \notag \\
&&A_{\bm k}=\frac{1}{\sqrt{2V\chi_{\perp}\omega_{\bm k}}}\ , \ \ \ \ \omega_{%
\bm k}=ck \ , \ \ \ \ c=\sqrt{\rho_s/\chi_{\perp}}\ .
\end{eqnarray}
Here ${\vec e}_{\lambda}$ ($\lambda=x,y$) is the magnon polarization, the
unit vector along the corresponding direction in the spin space; $a_{{\bm k}%
\lambda}^{\dag}$ is creation operator of the corresponding magnon; and $V$
is the area of the system.

Application of the Fermi golden rule to the Hamiltonian (\ref{nint}) with
account of (\ref{q1}) and the averaging over neutron polarizations gives the
elastic and inelastic scattering probabilities with momentum transfer ${\bm q%
}$ and energy transfer $\omega $ 
\begin{equation}
W(\omega ,{\bm q})=V\left[ \delta (\omega )\delta _{{\bm q},0}+2A_{\bm %
q}^{2}\delta (\omega -\omega _{\bm q})\right] \ .  \label{ww}
\end{equation}%
Here $\delta _{{\bm q},{\bm k}}$ is the Kronecker symbol. This gives the
well known result for the q-integrated scattering probability 
\begin{equation}
O(3):\ \ \ W_{0}(\omega )=\sum_{\bm q}W(\omega ,{\bm q})=V\left[ \delta
(\omega )+\frac{1}{2\pi \rho _{s}}\right] \ .  \label{wt}
\end{equation}%
The subscript 0 shows that this is the zero temperature case.

The O(2) nonlinear $\sigma $ model describes an easy-plane AF. The
Lagrangian is the same (\ref{L1}), but the spin vector is two dimensional, ${%
\vec{n}}=(n_{y},n_{z})$. Obviously, in this case the elastic scattering
probability is the same, and the inelastic one is twice smaller because
there is only one magnon. 
\begin{equation}
O(2):\ \ \ W_{0}(\omega )=\sum_{\bm q}W(\omega ,{\bm q})=V\left[ \delta
(\omega )+\frac{1}{4\pi \rho _{s}}\right] \ .  \label{wt1}
\end{equation}

\begin{figure}[h]
\centering \includegraphics[width=0.95
\columnwidth,clip=true]{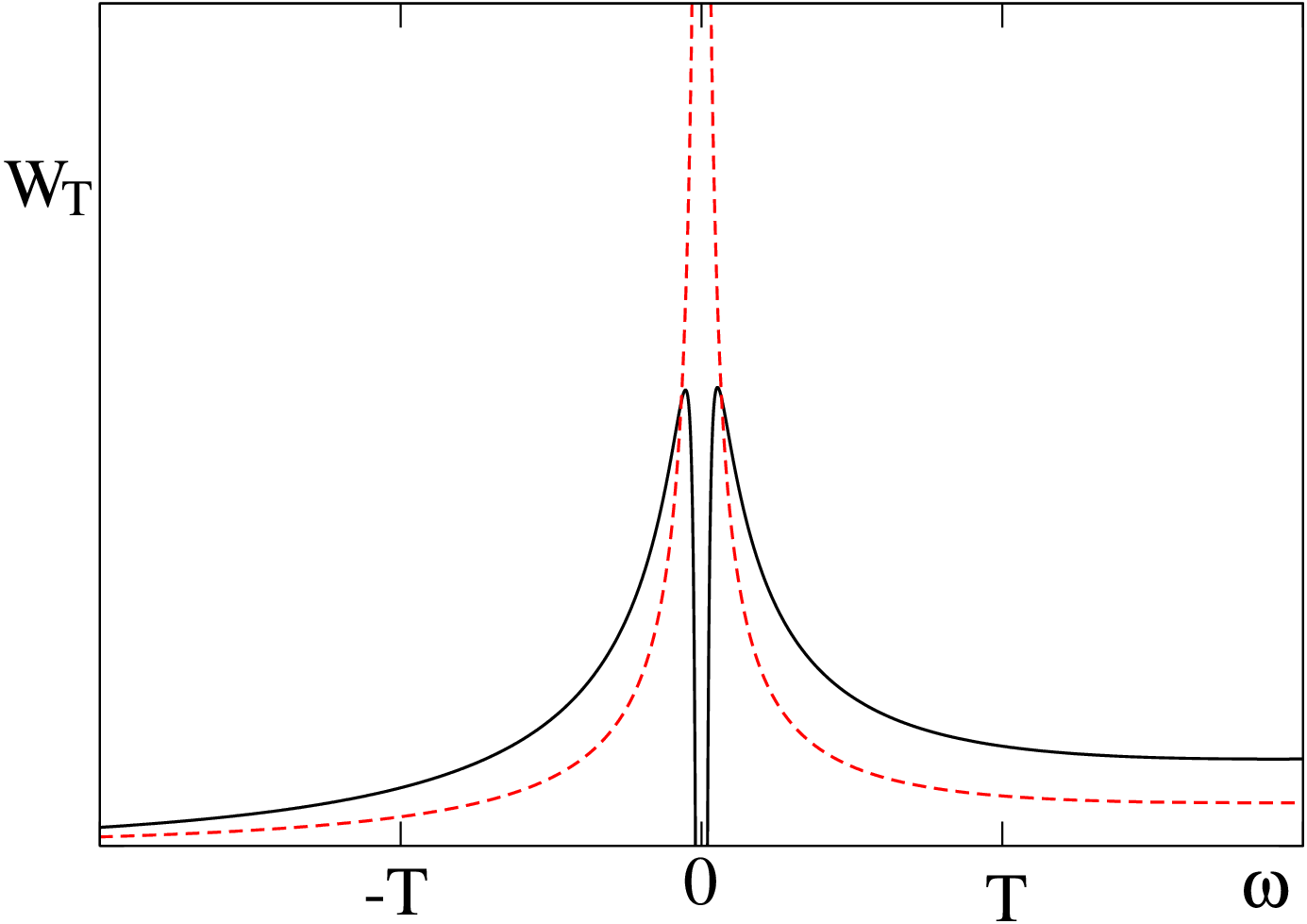}
\caption{\textit{(Color online)} A schematic sketch of the momentum
integrated neutron scattering probability $W_{T}(\protect\omega )$ at
nonzero temperature. The probability for an isotropic magnet (O(3) $\protect%
\sigma $-model) is shown by the black solid line. The probability for an
easy-plane magnet (O(2) $\protect\sigma $-model) is shown by the red dashed
line. }
\label{WT}
\end{figure}

It is well known that any small but finite temperature destroys the magnetic
ordering, so the $\delta $-function peaks in both Eqs. (\ref{wt}) and (\ref%
{wt1}) disappear, they are replaced by broad quasielastic peaks. In the O(3)
case a nonzero temperature leads to a finite correlation length $\xi $ and
to the corresponding \textquotedblleft spin-wave gap\textquotedblright ~\cite%
{Chak} 
\begin{eqnarray}
&&\xi \sim \exp (1/\tau )  \notag \\
&&\Delta \sim c/\xi \sim 2\pi \rho _{s}e^{-1/\tau }\ .  \label{dd}
\end{eqnarray}%
We have set the lattice spacing equal to unity and introduced the
dimensionless temperature 
\begin{equation}
\tau =\frac{T}{2\pi \rho _{s}}\ll 1\ .  \label{aa}
\end{equation}%
In the O(3) case our result for the q-integrated neutron scattering
probability at nonzero temperature reads 
\begin{eqnarray}
&&O(3):  \label{WO3} \\
&&\frac{W_{T}(\omega )d\omega }{V}=\left\{ 
\begin{array}{ll}
\sim 0 & \ \ \ \ |\omega |\lesssim \Delta \\ 
\tau \frac{d\omega }{|\omega |}\left( 1-\frac{\ln (T/|\omega |)}{\ln
(T/\Delta )}\right) & \ \ \ \ \Delta \lesssim |\omega |\lesssim T \\ 
\sim 0 & \ \ \ \ \omega \lesssim -T \\ 
\frac{d\omega }{2\pi \rho _{s}} & \ \ \ \ \omega \gtrsim T%
\end{array}%
\right.  \notag
\end{eqnarray}%
The probability is sketched in Fig.\ref{WT} by the black solid line. 
The nontrivial point is that the quasielastic peak at $|\omega |\lesssim T$
differs from the previously known formula~\cite{AA} by the logarithmic term $%
-\tau d\omega /|\omega |\ln (T/|\omega |)/\ln (T/\Delta )$. The term is
important to satisfy the spectral sum rule: the total quasielastic intensity
at nonzero temperature must be equal to the elastic intensity at zero
temperature 
\begin{equation}
\int_{\sim -T}^{\sim T}W_{T}(\omega )d\omega \approx \int_{-\delta
\rightarrow 0}^{+\delta \rightarrow 0}W_{0}(\omega )d\omega =V\ .
\label{srule}
\end{equation}%
This sum rule must be valid with relative accuracy up to terms of the order
of $\sim \tau \ll 1$. Without account of the logarithmic term the sum rule
is violated. It is easy to see that the term reduces the total quasielastic
intensity by two times. A simple heuristic derivation of the spectrum (\ref
{WO3}) is given in section IV, and a rigorous derivation based on the RG
analysis is given in section V.

An easy-plane quantum magnet is described by the O(2) nonlinear $\sigma$%
-model. The long range magnetic order is also destroyed by a nonzero
temperature, but in this case there is no a correlation length and no a
``spin-wave gap''. Because of the Berezinsky-Kosterlitz-Thouless physics all
correlators decay as powers of distance~\cite{Chak}. Our result for the
nonzero temperature q-integrated neutron scattering probability in this case
reads 
\begin{eqnarray}  \label{WO2}
&&O(2): \\
&&\frac{W_T(\omega)d\omega}{V}= \left\{ 
\begin{array}{ll}
\frac{1}{2}\tau T^{-\tau}\frac{d\omega}{|\omega|^{1-\tau}} & \ \ \ \
|\omega| \lesssim T \\ 
\sim 0 & \ \ \ \ \omega \lesssim -T \\ 
\frac{d\omega}{4\pi\rho_s} & \ \ \ \ \omega \gtrsim T%
\end{array}%
\right.  \notag
\end{eqnarray}
The probability is sketched in Fig.\ref{WT} by the red dashed line. The
energy scale $\Delta$ plays a role in this case as well. Certainly it is not
a gap any more, this is just the relevant energy scale. At $|\omega| \gg
\Delta$ the intensity (\ref{WO2}) is twice smaller than (\ref{WO3}). This is
because there is only one magnon in the O(2) case against two magnons in the
O(3) case. On the other hand at $|\omega| \lesssim \Delta$ the O(2)
intensity is much larger than the O(3) one. The spectral sum rule (\ref%
{srule}) is satisfied in this case as well because the elastic intensity is
the same in both O(2) and O(3) cases. A heuristic derivation of the spectrum
(\ref{WO2}) is given in section IV, and a rigorous RG derivation is given
in section V.

\begin{figure}[h]
\centering \includegraphics[width=0.95\columnwidth,clip=true]{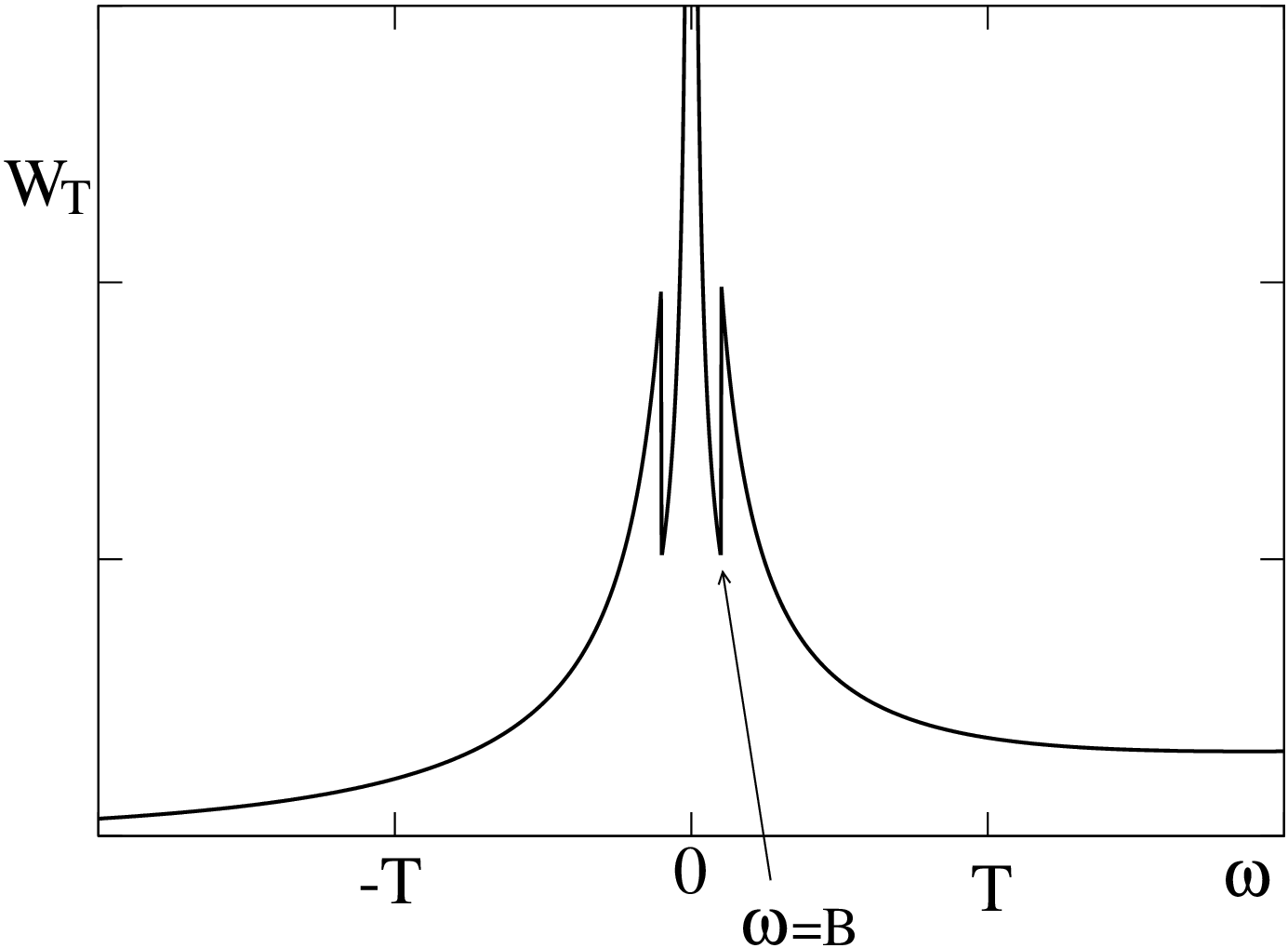}
\caption{A schematic sketch of the momentum integrated neutron scattering
probability $W_{T}(\protect\omega )$ for isotropic antiferromagnet (O(3) $%
\protect\sigma $-model) in the external magnetic field in the regime $\Delta
\ll B\ll T$. }
\label{WBT1}
\end{figure}

A uniform magnetic field, applied to an isotropic antiferromagnet, orients
sublattice moments perpendicular to the field and therefore acts as a kind
of easy plane anisotropy. This means that the magnetic field smoothly
transfers the O(3) magnet to the O(2) magnet. If temperature is zero then
the effect of the magnetic field acting on isotropic antiferromagnet is very
simple and well understood. We consider here a nonzero temperature. For
convenience we rescale the magnetic field, $g\mu _{B}B\rightarrow B$, where $%
g$ is the g-factor, and $\mu _{B}$ is Bohr magneton. If the magnetic field
is larger than temperature, $B>T$, then obviously the quasielastic peak is
described by the O(2) formula (\ref{WO2}). If the magnetic field is very
small, $B\ll \Delta $, then practically the quasielastic peak is described
by the O(3) formula (\ref{WO3}). The only nontrivial case is the case of the
intermediate magnetic field, $\Delta \ll B\ll T$. Our result for the
q-integrated spectrum in this case reads 
\begin{eqnarray}
&&O(3)\ \text{in\ magnetic\ field},\ \ \Delta \ll B\ll T:  \label{WBT} \\
&&\hspace{-10pt}\frac{W_{T}(\omega )d\omega }{V}=\left\{ 
\begin{array}{ll}
\frac{\tau }{2}\frac{B^{-\tau ^{\ast }}d\omega }{|\omega |^{1-\tau ^{\ast }}}%
\left[ 1-\tau \ln \left( \frac{T}{B}\right) \right]  & \ \ \ |\omega
|\lesssim B \\ 
\tau \frac{d\omega }{|\omega |}\left[ 1-\tau \ln \left( \frac{T}{|\omega |}%
\right) \right]  & \ \ \ B\lesssim |\omega |<T \\ 
\sim 0 & \ \ \ \omega \lesssim -T \\ 
\frac{d\omega }{2\pi \rho _{s}} & \ \ \ \omega \gtrsim T%
\end{array}%
\right.   \notag
\end{eqnarray}%
where 
\begin{equation}
\tau ^{\ast }=\frac{\tau }{1-\tau \ln \left( T/B\right) }\ .  \label{t*}
\end{equation}%
The probability (\ref{WBT}) is sketched in Fig.\ref{WBT1}. Naturally, the
spectrum (\ref{WBT}) satisfies the spectral sum rule (\ref{srule}). 
A heuristic derivation of the spectrum (\ref{WBT}) is given in section IV,
and a rigorous RG derivation is given in section V.

\section{Dependence of elastic/quasielastic neutron scattering intensity on
magnetic field in a vicinity of a magnetic O(3) QCP}

In this section we discuss enhancement of the staggered magnetization by
applied external magnetic field. This effect was first pointed out for the
square lattice Heisenberg antiferromagnet in Ref.\onlinecite{zhitomirsky98}.
In the
present paper we stress that the effect is strongly enhanced in the vicinity
of a magnetic QCP. To discuss this effect it is sufficient to consider the
zero temperature case. At a finite temperature the integrated quasielastic
intensity equals to the elastic intensity at zero temperature. The magnetic
field changes the staggered magnetization, 
\begin{eqnarray}
\langle n_{z}\rangle &\approx &1-\frac{1}{2}\langle n_{\perp }^{2}\rangle +%
\frac{1}{2}\langle n_{\perp }^{2}\rangle _{B=0}  \label{q2} \\
&=&1-\frac{1}{2V\chi _{\perp }}\sum_{\bm k}\left( \frac{1}{\omega _{\bm k}}-%
\frac{1}{\omega _{\bm k,B=0}}\right) \ .  \notag
\end{eqnarray}%
We remind that our normalization is $\langle n_{z}\rangle _{T=0,B=0}=1$. In
the presence of external magnetic field the in-plane magnon remains gapless
(easy plane), $\omega _{\bm k}=ck$, while the out-of-plane magnon becomes
gapped~\cite{com}, $\omega _{\bm k}=\sqrt{c^{2}k^{2}+B^{2}}$. This results
in the enhancement of the staggered magnetization, 
\begin{equation}
\delta \langle n_{z}\rangle =\frac{1}{4\chi _{\perp }}\int \frac{d^{2}k}{%
(2\pi )^{2}}\left( \frac{1}{ck}-\frac{1}{\sqrt{c^{2}k^{2}+B^{2}}}\right) =%
\frac{|B|}{8\pi \rho _{s}}\ .  \label{dn}
\end{equation}
Clearly this formula is valid only if $\delta \langle n_{z}\rangle \ll 1$.
There are two important points to note about Eq.(\ref{dn}): (i) the
dependence on the magnetic field is linear, (ii) the dependence is enhanced
approaching a QCP where the spin stiffness $\rho _{s}$ vanishes (The Eq. (%
\ref{dn}) is however inapplicable in the near vicinity of the QCP where the
criterion {$\delta \langle n_{z}\rangle \ll 1$ is violated}),
see e.g. Ref.~\onlinecite{Sachdev}.

To estimate magnitude of the effect (\ref{dn}) we refer to the simple model~%
\cite{HSS}, two coupled Heisenberg planes with spin 1/2, the position of the
QCP is $g_{c}=J_{\perp }/J=2.525$. In the vicinity of the QCP the {observed
staggered magnetization $\mu $ and the spin stiffness scale as $\mu \propto
(g_{c}-g)^{\beta }$, $\rho _{s}\propto (g_{c}-g)^{\nu }$} , where the
critical indexes are related, $2\beta =(1+\eta )\nu $, see e.g. 
Refs.~\onlinecite{Chak,Chub}. 
The index $\eta \approx 0.03$ is very small, so practically $%
\nu \approx 2\beta $. The spin stiffness at $J_{\perp }=0$ is $\rho
_{s0}\approx 0.2J$ and the staggered magnetic moment $\mu ^{2D}\approx
0.6\mu _{B}$, see Ref.~\onlinecite{SZ}. Hence

\begin{equation}  \label{estr}
\rho_s \sim 0.2J\left(\frac{\mu}{0.6\mu_B}\right)^2 \ .
\end{equation}

\begin{figure}[h]
\centering \includegraphics[width=0.9\columnwidth,clip=true]{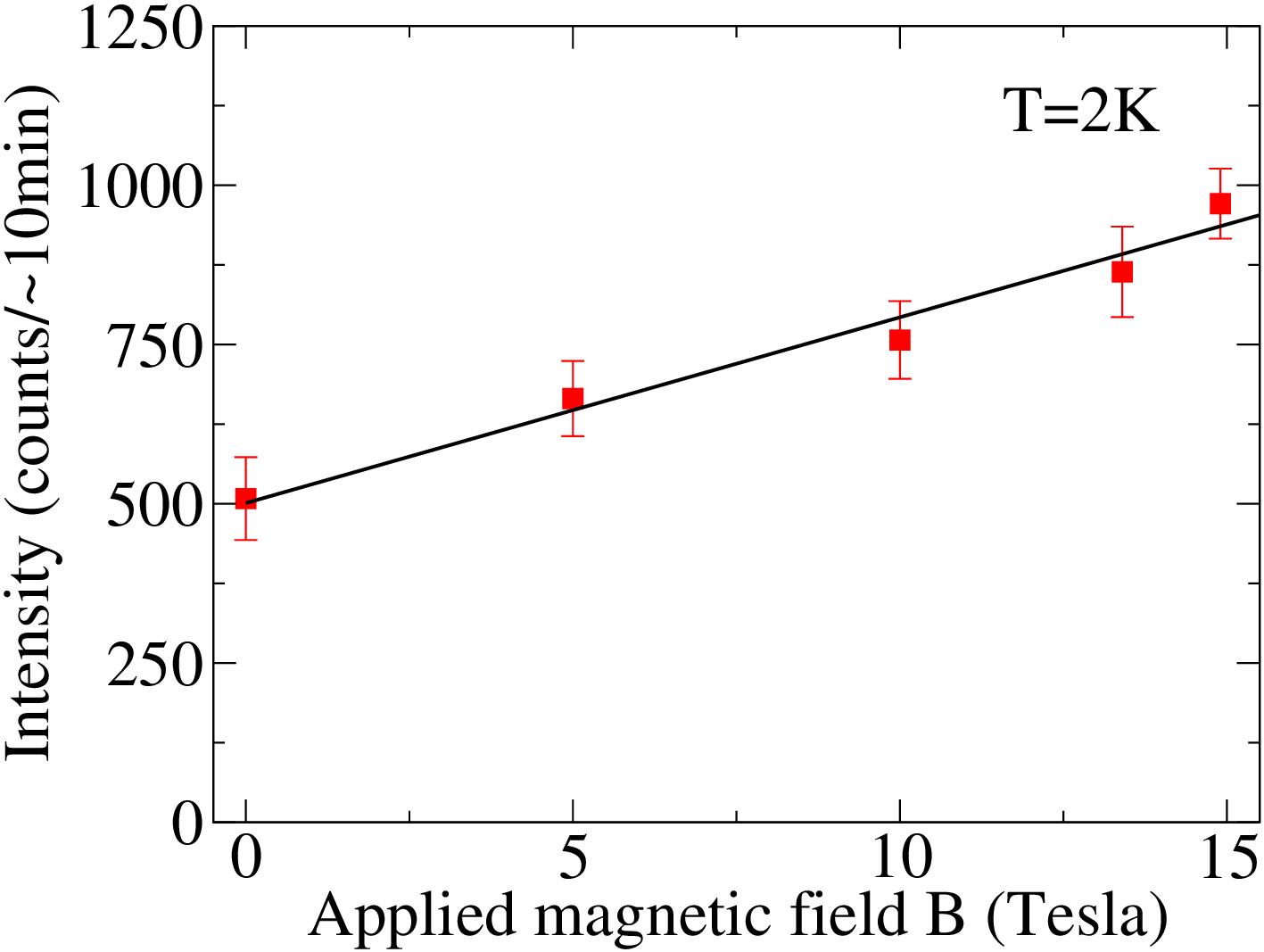}
\caption{\textit{(Color online)}
YBa$_2$Cu$_3$O$_{6.45}$: The experimental dependence~\protect\cite%
{Haug09} of the elastic peak intensity on the external magnetic field at $%
T=2 $~K. The intensity is measured at the incommensurate position of the
neutron scattering peak. The line is the result of a linear fit to the data
points. }
\label{FieldYBCO}
\end{figure}

For further numerical estimates we refer to the experiment~\cite{Haug09}
performed with YBa$_{2}$Cu$_{3}$O$_{6.45}$. The experimental dependence of
the quasielastic neutron scattering intensity on magnetic field is shown in
Fig.\ref{FieldYBCO} 
Note that the compound is a superconductor with critical superconducting
temperature $T_c=35K$. This is the second order superconductor, the magnetic
field of several Tesla pretty much penetrates in the bulk, and therefore we
disregard superconductivity in the analysis. The compound is very close to
the magnetic QCP, the value of the static ``staggered'' moment is $\mu \sim
0.05\mu_B$, see Ref.~\onlinecite{Haug09}. 
Therefore, according to Eq.(\ref{estr})
the spin stiffness is about $\rho _{s}\sim 2\times 10^{-3}J$. With the
magnetic field $B=15$ Tesla and with the characteristic cuprate exchange
integral $J=1400K$, Eq. (\ref{dn}) gives the following field induced
variation of the magnetization 
\begin{equation}
\frac{\delta \mu }{\mu }=\frac{|B|}{8\pi \rho _{s}}\sim 0.3\ .  \label{dn1}
\end{equation}
This value is very close to the data~\cite{Haug09}
 presented in
Fig.\ref{FieldYBCO} (Intensity is quadratic in magnetic moment, therefore $%
\delta I/{I }=2\delta \mu /{\mu }$). There is no doubt that the magnetic QCP
observed in YBa$_{2}$Cu$_{3}$O$_{y}$ is of a more complex nature than just a
simple O(3) QCP, see a discussion in Ref.~\onlinecite{Milstein08}.
 Nevertheless, it
is clear that generically Eq.(\ref{dn}) must be valid because it is based
only on two physical inputs: gapless spin-wave spectrum and reduction of the
effective spin stiffness in the vicinity of the QCP. 
\begin{figure}[h]
\centering \includegraphics[width=0.9\columnwidth,clip=true]{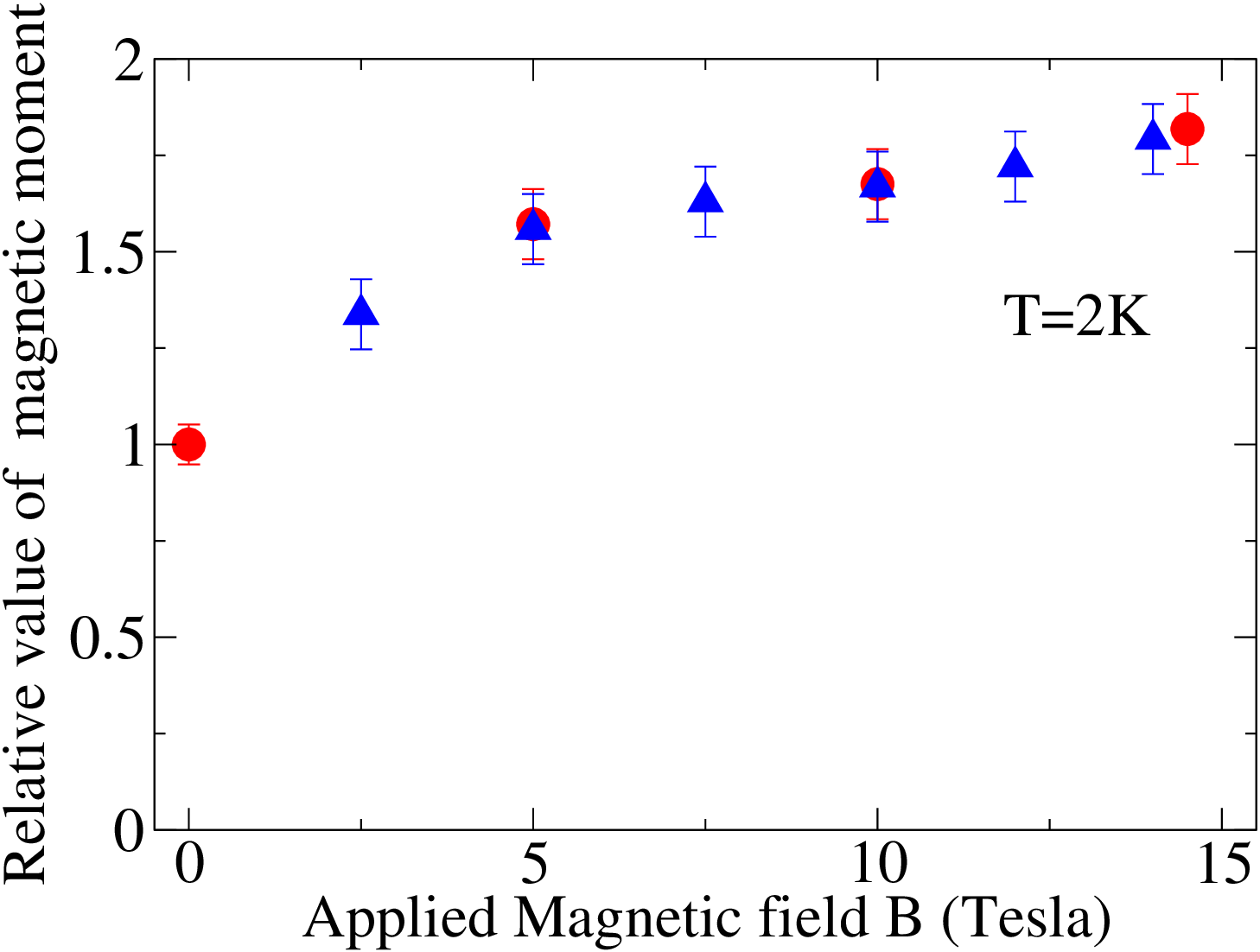}
\caption{\textit{(Color online)}
La$_{1.9}$Sr$_{0.1}$CuO$_4$: The experimental dependence~%
\protect\cite{Lake02} of the static magnetic moment on the external magnetic
field at $T=2$~K. The magnetic moment is measured at the incommensurate
position of the neutron scattering peak. The data points shown by red
circles are obtained from Fig.3a of Ref.~\onlinecite{Lake02}, and the
data points shown by blue triangles are obtained from Fig.3b of 
Ref.~\onlinecite{Lake02}. }
\label{FieldLSCO}
\end{figure}

The La based compound La$_{2-x}$Sr$_{x}$CuO$_{4}$ is very strongly
disordered due to random positions of Sr ions. In addition the compound has
a nonmonotonous dependence of the static magnetic moment on doping~\cite%
{muofx} (anomaly at $x\approx 0.12$). These two reasons complicate a
comparison between theory and experiment. The magnetic field dependence of
the static magnetic moment in La$_{1.9}$Sr$_{0.1}$CuO$_{4}$ has been
measured by neutron diffraction.~\cite{Lake02} The experimental
magnetic moment versus magnetic field is plotted in Fig.\ref{FieldLSCO}. The
magnetic moment is normalized to its value at zero field. Fig.\ref{FieldLSCO}
displays the quick growth from zero to 5 Tesla and then a slow linear
dependence. A possible explanation of these data is based on the idea of
magnetism induced around superconducting vortex core.~\cite{DSZ} An
alternative explanation based on the approach developed in the present paper
is that the steep dependence up to 5 Tesla is due to glassy effects related
to random positions of Sr ions, and the slow dependence for $5<B<15$ is due
to suppression of quantum fluctuations described by Eq.(\ref{dn}). Within
this framework the effect observed in both YBa$_{2}$Cu$_{3}$O$_{6.45}$ and La%
$_{1.9}$Sr$_{0.1}$CuO$_{4}$ has the same physical origin.

\section{Heuristic derivation of inelastic neutron scattering spectra at
nonzero temperature}

We start from derivation of Eq.(\ref{WO3}) that describes quasielastic
scattering from the isotropic antiferromagnet (the O(3) case). Let us
consider the energy transfer $\omega$ much larger than the ``spin wave gap''
and somewhat smaller than temperature, $\Delta \ll \omega \lesssim T$. Hence
the momentum transfer is $q \sim \omega/c$. The corresponding time/space
scales, $\delta t \sim 1/\omega$, $\delta r \sim 1/q$, are much smaller than
the time/space scales of slow thermal fluctuations that destroy the
long-range AF order, $\delta t_f \sim 1/\Delta$, $\delta r_f \sim \xi$.
Thus, the neutron makes a snapshot of the slowly fluctuating
antiferromagnet, the AF order is untouched at the relevant scale. Hence, the
formula (\ref{ww}) for the scattering probability is almost correct. One
needs only to ``erase'' the elastic peak, and in the inelastic part to
account for thermally excited magnons with the standard Bose occupation
numbers 
\begin{equation}  \label{nk}
n_{\mathbf{k}}=\frac{1}{e^{\omega_{\mathbf{k}}/T}-1} \ .
\end{equation}
Hence, (\ref{ww}) is transformed to 
\begin{eqnarray}  \label{wwT}
W(\omega,{\bm q})= 2V A_{\bm q}^2\left[(n_{\mathbf{q}}+1)
\delta(\omega-\omega_{\bm q}) + n_{\mathbf{q}} \delta(\omega+\omega_{\bm q})%
\right] \ 
\end{eqnarray}
As usually the first term comes from the magnon emission and the second term
comes from the magnon absorption. Integrating (\ref{wwT}) over $q$ and
expanding at $|\omega| \ll T$ one finds 
\begin{equation}  \label{wtr}
W_T(\omega)d\omega\to \tau V \frac{d\omega}{|\omega|}\ .
\end{equation}
To propagate the spectrum down to $\omega \sim \Delta$ we note that the
power of $\omega$ is fixed by dimension. However, dimension does not forbid
an additional logarithmic contribution, $W_T\to W_T\left[1+ \mu
\ln(T/|\omega|)\right]$, where $\mu \ll 1$ is some small coefficient. The
logarithmic term must be negligible at $T\gtrsim |\omega| \gg \Delta$, where
Eqs.(\ref{wwT}),(\ref{wtr}) have been derived, however it can be 100\%
important at $|\omega| \sim \Delta$. Imposing the spectral sum rule
condition (\ref{srule}) one immediately finds that $\mu\approx-1/\ln(T/%
\Delta)\approx-\tau$. This concludes the heuristic derivation of the
spectrum (\ref{WO3}). Notice that the spectral sum rule uniquely determines
the spectrum as soon as one takes the first power of logarithm. However, a
priory, one cannot exclude higher powers, for example, $W_T\to W_T\left[1+
\mu \ln(T/|\omega|) + \nu \ln^2(T/|\omega|)\right]$. Therefore, the
presented simple heuristic derivation certainly cannot substitute a rigorous
derivation. The rigorous derivation of (\ref{WO3}) based on RG is given in
Section V.

In the O(2) case (easy-plane magnet) the logic of derivation is very
similar. In this case Eqs. (\ref{wwT}) and (\ref{wtr}) must be also valid in
the regime $\Delta \ll \omega \lesssim T$. The only difference is that the
prefactor in these Eqs. must be two times smaller because in the O(2) model
there is only one magnon instead of two magnons in the O(3) model. However,
propagation of these Eqs. to smaller $\omega $ is different because of the
Berezinsky-Kosterlitz-Thouless physics in the O(2) case. In this case there
is no a correlation length, all correlators decay as powers of appropriate
variables~\cite{Chak,Chub}. Therefore, the only possible correction to the
spectrum (\ref{wtr}) is an additional prefactor $(T/\omega )^{p}$, where $%
p\ll 1$. The quasielastic spectrum still has to satisfy the spectral sum
rule (\ref{srule}). From here one immediately finds that $p=-\tau $ and this
proves the validity of the spectrum (\ref{WO2}). An RG derivation of the
spectrum (\ref{WO2}) is presented in the Section V.

Eq. (\ref{WBT}) describes the quasielastic scattering from the isotropic
antiferromagnet in the external magnetic field, $\Delta \ll B\ll T$.
Obviously, at a frequency higher than the magnetic field, $B<|\omega |<T$,
the field is not important and hence the spectrum (\ref{WBT}) must coincide
with (\ref{WO3}). At $|\omega |<B$ the gapped magnon is switched off, the
problem is effectively reduced to the O(2) case, and hence the shape of the
spectrum must be the same as that in (\ref{WO2}). However, one has to take
care about thermal fluctuations with typical frequencies between $B$ and $T$%
. These fluctuations reduce value of the magnetization that comes to the
effective O(2) $\sigma $-model regime, 
\begin{equation}
\langle n_{z}\rangle ^{2}=1\rightarrow \langle n_{z}\rangle ^{2}=1-\langle
n_{\perp }^{2}\rangle =1-\tau \ln \left( \frac{T}{B}\right) \ .  \label{nzz}
\end{equation}
Neutron scattering is proportional to the magnetization squared, therefore, (%
\ref{nzz}) presents another explanation of Eq.(\ref{WO3}). Because of the
reduction of number of active magnons, $2\rightarrow 1$, the intensity at $%
\omega =B-0$ must jump down by factor 2 compared to the intensity at $\omega
=B+0$. Further evolution down in frequency is governed by the O(2) $\sigma $%
-model. However, we have defined the effective O(2) $\sigma $-model with $%
\langle n_{z}\rangle =1$ at the upper cutoff which is $B$ in this case
instead of $T$ in (\ref{WO2}). Therefore to apply (\ref{WO2}) one has to
renormalize magnetization back to unity, $\langle n_{z}\rangle \rightarrow 1$%
, and this implies that $\rho _{s}\rightarrow \rho _{s}^{\ast }=\rho _{s}%
\left[ 1-\tau \ln \left( \frac{T}{B}\right) \right] $. This explains why $%
\tau $ defined in (\ref{aa}) is replaced by $\tau ^{\ast }$ given by (\ref%
{t*}). Notice that the spectrum (\ref{WBT}) satisfies the spectral sum rule (%
\ref{srule}). Rigorous RG derivation of this spectrum is presented in the
Section V.

\section{Renormalization Group (RG) analysis}

To perform calculations of thermodynamic and dynamic properties we consider
contribution of different momenta in small steps. To account for thermal
fluctuations, we put ultraviolet cutoff $\Lambda $ on the theory (which is
temperature $T/c\rightarrow T$ in the finite-temperature
renormalized-classical regime) and follow the standard procedure\cite%
{Polyakov,Nelson}. Supposing that magnetic field acts along the $x$-axis and
the order parameter is along the $z$-axis, we represent the vector $\mathbf{n%
}=(\pi _{x},\pi _{y},\sqrt{1-\pi ^{2}})$ and introduce the two-component
vector $\mathbf{\pi =}(\pi _{x},\pi _{y})$. In terms of the two-component
field $\pi $ the Lagrangian becomes (now we use the imaginary time $t$, {and
set the spin-wave velocity equal to unity, $c=1$}) 
\begin{eqnarray}
\mathcal{L} &=&\frac{\rho _{s}}{2}\int dtd^{2}x[(\partial \pi )^{2}+B^{2}\pi
_{x}^{2}+h\pi ^{2}  \notag \\
&&+(\mathbf{\pi }\partial \mathbf{\pi })^{2}+\frac{h}{4}\pi ^{4}-4iB\pi
^{2}\partial _{t}\pi _{c}-B^{2}  \notag \\
&&-\frac{1}{\rho _{s}}(\pi ^{2}+\frac{1}{2}\pi ^{4})]  \label{LL}
\end{eqnarray}%
where $\partial =(\nabla ,\partial _{t}),$ and we have introduced the source
field $h$ for the longitudinal ($n_{z}$) spin component. The terms in the
first line of Eq. (\ref{LL}) correspond to the Lagrangian of the
noninteracting spin waves, the other terms are responsible for the spin-wave
interaction (the last term arises from the condition $n^{2}=1$).

Let consider the renormalization of parameters of the spin wave spectrum by
the interaction. To first order in $1/\rho _{s}$ we have 
\begin{eqnarray}
\pi ^{4} &\rightarrow &2\pi ^{2}\sum\nolimits_{i=x,y}\langle \pi
_{i}^{2}\rangle +4\sum\nolimits_{i=x,y}\langle \pi _{i}^{2}\rangle \pi
_{i}^{2}  \notag  \label{Decoupl} \\
&&-\langle \pi ^{2}\rangle \langle \pi ^{2}\rangle
-2\sum\nolimits_{i=x,y}\langle \pi _{i}^{2}\rangle \langle \pi
_{i}^{2}\rangle  \notag \\
(\pi _{i}\partial \pi _{i})^{2} &\rightarrow &\pi _{i}^{2}\langle (\partial
\pi _{i})^{2}\rangle +(\partial \pi _{i})^{2}\langle \pi _{i}^{2}\rangle 
\notag \\
&&-\langle \pi _{i}^{2}\rangle \langle (\partial \pi _{i})^{2}\rangle
\end{eqnarray}%
After performing decouplings (\ref{Decoupl}) we obtain up to a constant
contribution 
\begin{eqnarray}
\mathcal{L} &=&\frac{{\rho _{s}}}{2}\int dtd^{2}x\left[ \sum%
\nolimits_{i=x,y}(\partial \pi _{i})^{2}(1+\langle \pi _{i}^{2}\rangle
)\right. \\
&&\left. +B^{2}\pi _{x}^{2}(1-\langle \pi _{x}^{2}\rangle )+h\pi ^{2}+2B\pi
^{2}\partial _{t}\pi _{y}+O(\pi ^{4})^{^{^{^{{}}}}}\right]  \notag
\end{eqnarray}%
where we have taken into account $\langle (\partial \pi _{i})^{2}\rangle -1/{%
\rho _{s}}=-m_{i}^{2}\langle \pi _{i}^{2}\rangle $; $m_{x}^{2}=B^{2}+h,$ $%
m_{y}^{2}=h$ are masses of $\pi _{x}$ and $\pi _{y}$ fields respectively.
Rescaling the field $\pi \rightarrow (1+\langle \pi _{i}^{2}\rangle
)^{-1/2}\pi $ to restore the coefficient at $(\partial \pi )^{2}$ we obtain%
\begin{eqnarray}
\mathcal{L} &=&\frac{{\rho _{s}}}{2}\int dtd^{2}x\left[ (\partial \pi
)^{2}+B^{2}\pi _{x}^{2}(1-2\langle \pi _{x}^{2}\rangle )\right.  \notag \\
&&+h\sum\nolimits_{i=x,y}\pi _{i}^{2}(1+\frac{1}{2}\langle \pi _{\overline{i}%
}^{2}\rangle -\frac{1}{2}\langle \pi _{i}^{2}\rangle )+  \notag \\
&&\left. +2B\pi ^{2}\partial _{t}\pi _{y}+O(\pi ^{4})^{^{^{{}}}}\right]
\end{eqnarray}%
where $\overline{x}=y$ and $\overline{y}=x.$ The corresponding renormalized
parameters are 
\begin{eqnarray}
B_{R}^{2} &=&B^{2}(1-2\langle \pi _{x}^{2}\rangle )  \notag \\
h_{R}^{i} &=&h(1+\frac{1}{2}\langle \pi _{\overline{i}}^{2}\rangle -\frac{1}{%
2}\langle \pi _{i}^{2}\rangle )
\end{eqnarray}%
From the performed rescaling we have in addition 
\begin{equation}
\rho _{is}^{R}Z_{i}^{2}={\rho _{s}}(1+\langle \pi _{i}^{2}\rangle )
\end{equation}%
where $Z_{i}$ is the rescaling factor of the field $n_{i}$. Requiring
similar to~\cite{Polyakov,Nelson} $h^{R}\rho _{si}^{R}=Z_{i}(h\rho
_{si}),$ we obtain 
\begin{equation}
Z_{i}=1+\frac{1}{2}\langle \pi _{\overline{i}}^{2}\rangle +\frac{1}{2}%
\langle \pi _{i}^{2}\rangle
\end{equation}%
Therefore%
\begin{eqnarray}
Z_{x} &=&Z_{y}=1+\frac{1}{2}\langle \pi _{x}^{2}\rangle +\frac{1}{2}\langle
\pi _{y}^{2}\rangle  \notag \\
\rho _{is}^{R} &=&{\rho _{s}}(1-\langle \pi _{\overline{i}}^{2}\rangle )
\label{ren}
\end{eqnarray}%
The renormalization factor for the $n_{z}$ field can be determined from the
sublattice magnetization 
\begin{equation}
\langle n_{z}\rangle =1-\frac{1}{2}\langle \pi _{x}^{2}\rangle -\frac{1}{2}%
\langle \pi _{y}^{2}\rangle
\end{equation}%
which implies $Z_{x}=Z_{y}=Z_{z}$. Similarly to the isotropic case~\cite%
{Nelson}, the obtained expressions for the renormalized parameters (\ref{ren}%
) can be used to formulate the renormalization group transformation. Below
we consider the renormalized classical regime, $T\gg \max (\Delta ,B)$.

As it is argued in Refs.~\onlinecite{Chak,Chub}, it is sufficient in the
renormalized classical regime to account only for classical (zero Matsubara
frequency) contributions to $\langle \pi _{i}^{2}\rangle $. Considering the
contribution of infinitely thin layer in momentum space, we obtain ($\tau
_{i}=T/(2\pi \rho _{s}^{i})$) 
\begin{eqnarray}
\frac{d\tau _{x}}{dl} &=&\tau _{x}\tau _{y},  \notag  \label{tac} \\
\frac{d\tau _{y}}{dl} &=&\frac{\tau _{x}\tau _{y}}{1+e^{2l}B^{2}},  \notag \\
\frac{d\ln Z^{2}}{dl} &=&\frac{\tau _{x}}{1+e^{2l}B^{2}}+\tau _{y},
\end{eqnarray}%
where $l=\ln (\Lambda /\mu )$ is the scaling parameter. The flow of the
magnetic field can be neglected in the one-loop approximation.

For $B=0$ Eqs. (\ref{tac}) are equivalent to standard one-loop
RG equations for 2D antiferromagnet \cite{Polyakov,Nelson}.
Eqs. (\ref{tac}) are similar to those for the easy-plane case studied
in Ref.~\onlinecite{IK1}, although differ from those due to presence of two
spin stiffnesses instead of one. The other way of writing first two 
Eqs. (\ref{tac}) is
\begin{eqnarray}
\frac{d\ln \tau _{x}}{dl} &=&\tau _{y},  \notag \\
\frac{d\ln \tau _{y}}{dl} &=&\frac{\tau _{x}}{1+e^{2l}B^{2}}\ .
\end{eqnarray}%
Together with the last Eq. (\ref{tac}) this implies that $Z^{2}(l)/[\tau
_{x}(l)\tau _{y}(l)]=1/\tau ^{2}$ is the invariant of the flow.

At $e^{2l}B_{a}^{2}(l)\ll 1$ [$O(3)$ regime] we have $\tau _{x}(l)\approx
\tau _{y}(l)$ such that%
\begin{equation}
\frac{d\tau _{x,y}}{dl}=\tau _{x,y}^{2}
\end{equation}%
and $\tau _{x,y}(l)/Z(l)=\tau ,$ which yields 
\begin{eqnarray}
\tau _{x}(l) &\approx &\tau _{y}(l)\approx \tau /(1-\tau l);  \notag
\label{O3} \\
Z(l) &\approx &1/(1-\tau l)
\end{eqnarray}%
At $e^{2l}B_{a}^{2}(l)\sim 1$ the crossover from $O(3)$ to $O(2)$ regime
occurs. At the crossover scale $l^{\ast }=\ln (\Lambda /B)$ we obtain $\tau
^{\ast }\equiv \tau _{x}^{\ast }\approx \tau _{y}^{\ast }\approx \tau
Z^{\ast }$ with 
\begin{equation}
Z^{\ast }\approx 1/[1-\tau \ln (\Lambda /B)]  \label{Zstar1}
\end{equation}%
For $e^{2l}B_{a}^{2}(l)\gg 1$ [$O(2)$ regime] the flow of $\tau _{c}(l)$ is
suppressed and we have $\tau _{y}(l)=\tau _{y}^{\infty }$=const,%
\begin{equation}
\frac{d\ln Z^{2}}{dl}=\frac{d\ln \tau _{x}}{dl}=\frac{\tau _{y}^{\infty }}{%
2\pi }  \notag
\end{equation}%
where $\tau _{y}^{\infty }\simeq \tau ^{\ast }$ up to logarithmic accuracy,
such that 
\begin{eqnarray}
\tau _{x}(l) &\approx &\tau ^{\ast }\exp [\tau ^{\ast }(l-l^{\ast })]  \notag
\\
Z^{2}(l) &\approx &Z^{\ast 2}\exp [\tau ^{\ast }(l-l^{\ast })]
\end{eqnarray}%
Representing $l=\ln (\Lambda /\mu )$ we obtain 
\begin{eqnarray}
\tau _{x}(\mu ) &\approx &\tau Z^{\ast }(B/\mu )^{1/\ln (B/\Delta )}  \notag
\label{O2b} \\
Z^{2}(\mu ) &\approx &Z^{\ast 2}(B/\mu )^{1/\ln (B/\Delta )}
\end{eqnarray}%
where $Z^{\ast }$ is given by Eq. (\ref{Zstar1}) and to one-loop order 
$\Delta =\Lambda \exp (-1/\tau )$ (Note that the two-loop corrections yield 
$\Delta $ given by Eq. (\ref{dd}{), see Ref~\onlinecite{Chak}).} This power-law
dependence corresponds to the Berezinsky-Kosterlitz-Thouless behavior of
correlation functions (see below).

Now we discuss the behavior of correlation functions important for
experimental data. From our scaling analysis we can obtain the transverse
Green functions 
\begin{eqnarray}
\langle \langle n_{\mathbf{q}}^{i}|n_{-\mathbf{q}}^{i}\rangle \rangle
_{i\omega } &\equiv &\int\nolimits_{0}^{1/T}dt\langle T[n_{\mathbf{q}%
}^{i}(t)n_{\mathbf{q}}^{i}(0)]\rangle e^{i\omega t} \\
&=&\langle \langle \pi _{\mathbf{q}}^{i}|\pi _{-\mathbf{q}}^{i}\rangle
\rangle _{i\omega }=\frac{2\pi \tau _{i\mathbf{q}}}{TZ_{i\mathbf{q}}^{2}}%
\frac{1}{\omega ^{2}+\omega _{i\mathbf{q}}^{2}}  \notag
\end{eqnarray}%
where  {$\omega _{i\mathbf{q}}=\sqrt{\mathbf{q}^{2}+m_{i}^{2}}$, }$%
m_{x}=B,$ $m_{y}=0,$ $\tau _{i\mathbf{q}},Z_{i\mathbf{q}}$ are obtained from
the result of the solution of Eqs. (\ref{tac}) by substitution $l_{i}=\ln
(T/\omega _{i\mathbf{q}})$. Therefore, we obtain the results for the
functions at {$|\mathbf{q}|\gg B$} in the renormalized classical regime%
\begin{eqnarray}
\langle \langle n_{\mathbf{q}}^{x}|n_{-\mathbf{q}}^{x}\rangle \rangle
_{i\omega } &=&\langle \langle n_{\mathbf{q}}^{y}|n_{-\mathbf{q}}^{y}\rangle
\rangle _{i\omega }=\frac{2\pi \tau _{x}}{TZ^{2}}\frac{1}{\mathbf{q}%
^{2}+\omega ^{2}}  \notag \\
&\simeq &\frac{1}{\rho _{s}}\left( 1-\tau \ln \frac{\Lambda }{|\mathbf{q|}}%
\right) \frac{1}{\mathbf{q}^{2}+\omega ^{2}}  \label{SO3}
\end{eqnarray}%
This result agrees with the result for the correlation function derived for
the $O(3)$ model by Chakravarty, Halperin and Nelson\cite{Chak}. On the
other hand, in the scaling limit $l\longrightarrow \infty $ we obtain from
Eq. (\ref{O2b}) {\ 
\begin{equation}
\frac{\tau _{y}(\mu )}{Z^{2}(\mu )}=\tau \left( 1-\tau \ln \frac{\Lambda }{B}%
\right) \left( \frac{\mu }{B}\right) ^{1/\ln (B/\Delta )}
\end{equation}%
} which implies the results for the functions at $|\mathbf{q}|\ll B${\ 
\begin{equation}
\langle \langle n_{\mathbf{q}}^{i}|n_{\mathbf{q}}^{i}\rangle \rangle
_{\omega }=\frac{1}{\rho _{s}}\left( 1-\tau \ln \frac{\Lambda }{B}\right)
\left\{ 
\begin{array}{cc}
\frac{1}{B^{2}} & i=x \\ 
\frac{(|\mathbf{q}|/B)^{\alpha }}{q^{2}} & i=y%
\end{array}%
\right\}  \label{SO2}
\end{equation}%
} with $\alpha =1/\ln (B/\Delta ).$ In this regime we obtain the power law
corrections to the Green functions, which are in agreement with the
Berezinsky-Kosterlitz-Thouless solution of the O(2) model. Note, that both
results, (\ref{SO3}) and (\ref{SO2}) fulfill the sumrule%
\begin{equation}
\sum_{\mathbf{q},\omega }\left[ \langle \langle n_{\mathbf{q}}^{x}|n_{%
\mathbf{q}}^{x}\rangle \rangle _{\omega }+\langle \langle n_{\mathbf{q}%
}^{y}|n_{\mathbf{q}}^{y}\rangle \rangle _{\omega }\right] =1  \label{SR}
\end{equation}%
which is actually required for the total spectral weight including the
longitudinal Green function. This shows that the contribution of the
longitudinal Green function is contained implicitly in the calculated
transverse Green functions, which seems to be the general property of the $%
1/S$ expansion\cite{Note}.

For the structure factor, given by imaginary parts of the Green functions at
the real axis, we obtain%
\begin{eqnarray}
{W}(\mathbf{q},\omega ) &=&\frac{n_{\omega }}{\pi }\sum\nolimits_{i=x,y}%
\mathrm{Im}\langle \langle n_{\mathbf{q}}^{i}|n_{-\mathbf{q}}^{i}\rangle
\rangle _{\omega } \\
&=&\frac{\pi c^{2}n_{\omega }}{T\omega }\sum\nolimits_{i=x,y}\frac{\tau _{i%
\mathbf{q}}}{Z_{i\mathbf{q}}^{2}}\left[ \delta (\omega -\omega _{i\mathbf{q}%
})+\delta (\omega +\omega _{i\mathbf{q}})\right]  \notag
\end{eqnarray}%
where $n_{\omega }$ is the Bose function. Integration over $\mathbf{q}$
yields%
\begin{eqnarray}
&&{W}(\omega )=\int \frac{d^{2}\mathbf{q}}{(2\pi )^{2}}{W}(\mathbf{q},\omega
)  \notag \\
&=&\frac{{\pi }}{\omega ^{2}}\int \frac{qdq}{2\pi }\sum_{i}\frac{\tau _{i%
\mathbf{q}}}{Z_{i\mathbf{q}}^{2}}\left[ \delta (\omega -\omega _{iq})+\delta
(\omega +\omega _{iq})\right]  \notag \\
&=&\frac{1}{2|\omega |}\sum_{i=x,y}\theta (\omega ^{2}-m_{i}^{2})\frac{\tau
_{i\omega }}{Z_{\omega }^{2}}
\end{eqnarray}%
where $\tau _{i\omega }=\tau _{i}(l=\ln (T/|\omega |))$ and similar for $%
Z_{\omega }$.

In the O(3) regime $|\omega |>B$ where $\tau _{x\omega }/Z_{\omega
}^{2}=\tau _{y\omega }/Z_{\omega }^{2}=\tau \lbrack 1-\tau {\ln (\Lambda
/|\omega |)}],$ we obtain 
\begin{equation}
{W}(\omega )=\frac{\tau }{|\omega |}\left( 1-\tau \ln \frac{\Lambda }{%
|\omega |}\right)
\end{equation}%
which agrees with the result (\ref{WO3}). In the O(2) regime ($|\omega |<B$) 
\begin{eqnarray}
\frac{\tau _{x\omega }}{Z_{\omega }^{2}} &=&\text{const}  \notag  \label{IO2}
\\
\frac{\tau _{y\omega }}{Z_{\omega }^{2}} &=&\tau \left( 1-\tau \ln \frac{%
\Lambda }{B}\right) \left( \frac{|\omega |}{B}\right) ^{\tau ^{\ast }}
\end{eqnarray}%
and%
\begin{equation}
{W}(\omega )=\frac{\tau }{2|\omega |}\left( 1-\tau \ln \frac{\Lambda }{B}%
\right) \left( \frac{|\omega |}{B}\right) ^{\tau ^{\ast }}
\end{equation}%
in agreement with (\ref{WBT}).

\section{Conclusion}

We have considered two dimensional quantum antiferromagnets. At zero
temperature the antiferromagnets possess the long range order that results
in the elastic Bragg peak observed in neutron scattering. At an arbitrary
small but finite temperature the long range order is destroyed and hence the
elastic Bragg peak is transformed to the quasielastic spectrum. We derive
these spectra. The derived quasielastic spectrum for an isotropic
antiferromagnet is given by Eq.(\ref{WO3}) and the corresponding physics is
determined by the finite correlation length. {The derived spectrum (\ref{WO3}%
) differs from the previously known answer~\cite{AA} by a logarithmic term
that changes the integrated intensity by two times.} The derived
quasielastic spectrum for an easy plane antiferromagnet is given by Eq.(\ref%
{WO2}) and the corresponding physics is determined by the
Berezinsky-Kosterlitz-Thouless power law. An external uniform magnetic field
drives a crossover from the isotropic antiferromagnet to the easy plane one.
Hence the quasielastic spectrum evolves with the magnetic field. This
evolution is described by Eq.(\ref{WBT}).

An external uniform magnetic field, applied to an isotropic quantum
antiferromagnet, suppresses quantum fluctuations. As a result the integrated
intensity of elastic (quasielastic) neutron scattering grows linearly with
applied external magnetic field, see Eq.(\ref{dn}). The dependence is
significantly enhanced in the vicinity of the quantum critical point that
separates magnetically ordered and magnetically disordered states, an
estimate is given in Eq.(\ref{dn1}).{\ We believe that the observed
enhancement of the quasielastic scattering in magnetic field in YBa$_{2}$Cu$%
_{3}$O$_{6.45}$, Ref.~\onlinecite{Haug09}, and in La$_{1.9}$Sr$_{0.1}$CuO$_{4}$,
Ref.~\onlinecite{Lake02}, is due to this mechanism. Extension of
the present analysis to include the magnetic incommensurability and
contribution of conduction electrons is of a great interest.

\section{Acknowledgements}

We are grateful to J. Oitmaa and G. S. Uhrig who attracted our attention to
the enhancement of staggered magnetization in magnetic field. We acknowledge
discussions with A. I. Milstein. A. K. acknowledges Max-Planck Society for
partial financial support within the Partnership Program. O.P.S acknowledges
MPI-PKS, Dresden, where part of this work was performed within the framework
of the Advanced Study Group "Unconventional Magnetism in High Fields".


\begin{thebibliography}{99}
\bibitem{com} Note that the formula $\omega_{\bm k}=\sqrt{c^2k^2+B^2}$ is
obvious without any calculations because projection of spin on magnetic
field is conserved and for the out-of plane magnon $\Delta S_B=1$.
\end{thebibliography}
\end{document}